\documentclass[aps,prx,preprint,onecolumn,citeautoscript,superscriptaddress,nofootinbib,
eqsecnum]{revtex4}
\usepackage{amsmath}
\usepackage{amssymb}
\usepackage{bbm}
\usepackage{bm}
\usepackage{comment}
\usepackage{graphicx}
\usepackage{color}
\usepackage[papersize={8.5in,11in}]{geometry}
\usepackage[colorlinks=true]{hyperref}
\hypersetup{
    bookmarks=true,         
    unicode=false,          
    pdftoolbar=true,        
    pdfmenubar=true,        
    pdffitwindow=false,     
    pdfstartview={FitH},    
    pdftitle={Spin glass and spin liquid co-existence in the random quantum Heisenberg magnet},    
    pdfauthor={M. Christos, F. M. Haehl, S. Sachdev},     
    pdfsubject={},   
    pdfcreator={},   
    pdfproducer={}, 
    pdfkeywords={} {} {}, 
    pdfnewwindow=true,      
    colorlinks=true,       
    linkcolor=magenta, 
    citecolor=blue,        
    filecolor=magenta,      
    urlcolor=blue           
}

\usepackage{amsfonts}
\usepackage{upgreek}
\usepackage{slashed}
\usepackage{latexsym}
\usepackage[export]{adjustbox}
\usepackage{dsfont}
\usepackage{subcaption}
\newcommand{\beq}{\begin{equation}}
\newcommand{\eeq}{\end{equation}}
\def\bea{\begin{eqnarray}}
\def\eea{\end{eqnarray}}

\begin{document}


\title{Spin liquid to spin glass crossover\\ in the random quantum Heisenberg magnet}

\author{Maine Christos}
\affiliation{Department of Physics, Harvard University, Cambridge MA 02138, USA}

\author{Felix M. Haehl}
\affiliation{School of Natural Sciences, Institute for Advanced Study, 1 Einstein Drive, Princeton, NJ 08540, USA}

\author{Subir Sachdev}
\affiliation{Department of Physics, Harvard University, Cambridge MA 02138, USA}
\affiliation{School of Natural Sciences, Institute for Advanced Study, 1 Einstein Drive, Princeton, NJ 08540, USA}

\date{\today}

\begin{abstract}
We study quantum SU($M$) spins with all-to-all and random Heisenberg exchange interactions of root-mean-square strength $J$.
The $M \rightarrow \infty$ model has a spin liquid ground state with the spinons obeying the equations of the Sachdev-Ye-Kitaev (SYK) model. Numerical studies of the SU(2) model with $S=1/2$ spins show spin glass order in the ground state, but also display SYK spin liquid behavior in the intermediate frequency spin spectrum. We employ a $1/M$ expansion to describe the crossover from fractionalized fermionic spinons
to a confining spin glass state with weak spin glass order $q_{EA}$.
The SYK spin liquid behavior persists down to a frequency $\omega_\ast \sim J q_{EA}$, and for $\omega < \omega_\ast$, the spectral density is linear in $\omega$, thus quenching the extensive zero temperature entropy of the spin liquid. The linear $\omega$ spectrum is qualitatively similar to that obtained earlier using bosonic spinons for large $q_{EA}$.
We argue that the extensive SYK spin liquid entropy is transformed as $T \rightarrow 0$ to an extensive complexity of the spin glass state.
\end{abstract}
\maketitle
\tableofcontents

\section{Introduction}
\label{sec:intro}

A common theme in many experimental studies of the hole-doped cuprate compounds below optimal doping is that while there is nearly static spin or charge order at low temperatures, the intermediate temperature pseudogap regime can be described in terms of an underlying spin liquid state. Among recent studies, we note the observations in La$_{2-x}$Sr$_x$CuO$_4$ of  Frachet {\it et al.\/} \cite{Julien19} showing spin glass order at low temperature all the way up to optimal doping; and of Fang {\it et al.} \cite{Ramshaw20} showing evidence for the breakdown of the Luttinger Fermi surface in the pseudogap, which can be interpreted in terms of a fractionalized Fermi liquid containing a background spin liquid \cite{ancilla1}.  
In the undoped antiferromagnet, we recall the observations of Dalla Piazza {\it et al.\/} \cite{Ronnow15} showing intermediate energy spinon continua at wavevector $(\pi, 0)$ in a system with long-range N\'eel order at wavevector $(\pi, \pi)$.

In this paper, we will study a random quantum Heisenbeg magnet with all-to-all exchange interactions $J_{ij}$
\beq
H = \frac{1}{\sqrt{N}} \sum_{i<j=1}^N J_{ij} {\bm S}_i \cdot {\bm S}_j\,. \label{SU2Ham}
\eeq
We study an ensemble of models, where the $J_{ij}$ are independent random variables for each pair $(i,j)$, and their ensemble averages are
\beq
\overline{J_{ij}} = 0 \quad, \quad \overline{J_{ij}^2} = J^2\,.
\eeq
This model generalizes the classical Sherrington-Kirkpatrik model with Ising spins $\sigma_i = \pm 1$ to quantum SU(2) spins ${\bm S}_i$, acting on a Hilbert space of states with angular momentum $S=1/2$ on each site. 

Although such a random exchange model is far from the microscopic situation in the cuprates, it can successfully capture many aspects of cuprate phenomenology \cite{CGPS}. Here, we will show that it exhibits a deconfinement-to-confinement crossover, and we will obtain explicit results for the dynamic spin susceptibility across this crossover. This is one of the rare instances in which a loss of fractionalization can be described in a strongly-coupled system with gapless matter.

The generalization of the model (\ref{SU2Ham}) to SU($M$) spins, and the limits $N \rightarrow \infty$ followed by $M \rightarrow \infty$, yield a fractionalized spin liquid ground state \cite{SY92} whose fermionic spinons obey the same equations as the complex Sachdev-Ye-Kitaev (SYK) model \cite{kitaev2015talk,Sachdev:2015efa,Gu2019,Tikhanovskaya:2020elb}. On the other hand, numerical studies \cite{GrempelRozenberg98,ArracheaRozenbergSG2002,Shackleton2021} of the $N \rightarrow \infty$ limit of the model (\ref{SU2Ham}) for SU(2) and spin $S=1/2$ show the presence of spin glass order in the ground state (in contrast to the SYK model itself, which does not have spin glass order \cite{Gur-Ari:2018okm}). However, the recent numerical study of the $S=1/2$ SU(2) model argued \cite{Shackleton2021} that the spin spectral density at intermediate frequencies matched that of the SYK spin liquid. Specifically, they observed 
\beq
\chi'' (\omega) \sim  \frac{\mbox{sgn}(\omega)}{J} \left[ 1 - \frac{c}{J} |\omega| \ldots \right] \quad , \quad \omega_\ast < |\omega| \ll J\,, T=0\,. \label{chisgn}
\eeq
The leading term in (\ref{chisgn}) has its origins in the spinons obeying the SYK equations \cite{SY92}; it is often called the `marginal' spectrum, because electrons scattering off such spin fluctuations acquire a marginal Fermi liquid Green's function. (The subleading term, with positive co-efficient $c$, is related to the co-efficient of the Schwarzian effective action \cite{Tikhanovskaya:2020elb}.) A similar marginal spectrum is obtained in the density fluctuations in a model with density-density interactions, and this has been argued \cite{Joshi20} to be related to the anomalous continuum observed in dynamic charge response measurements \cite{Abbamonte1,Abbamonte2}
on optimally doped Bi$_{2.1}$Sr$_{1.9}$Ca$_{1.0}$Cu$_{2.0}$O$_{8+x}$ (Bi-2212) using momentum-resolved electron energy-loss spectroscopy (M-EELS). 
In the present paper, we will obtain an estimate of the low frequency bound $\omega_\ast$ of the marginal spectrum, and also describe the nature of the crossover at $\omega \sim \omega_\ast$.

This paper addresses the nature of the crossover from the spectrum in (\ref{chisgn}) to frequencies $\omega < \omega_\ast$. The presence of spin glass order implies a delta function at zero frequency
\beq
\frac{T \chi'' (\omega)}{\omega} =  \pi q_{EA} \, \delta(\omega) \label{chidelta}
\eeq
where $q_{EA}$ is the spin glass order parameter. We find that the crossover occurs at a frequency
\beq
\omega_\ast = J q_{EA}\,, \quad \mbox{for $q_{EA} \ll 1$}, \label{omegastar}
\eeq
and for smaller frequency
\beq
\chi'' (\omega) \sim  \frac{\omega}{\omega_\ast \pi J}\quad , \quad 0< |\omega| < \omega_\ast\,, \; T=0\,. \label{chilinear}
\eeq
Given the numerical estimate $q_{EA} \sim 0.02$ \cite{Shackleton2021}, the spin liquid behavior of (\ref{chisgn}) is visible over a wide range of frequencies. 

We note that a linear spectrum, qualitatively similar to (\ref{chilinear}), was found in an earlier theory \cite{GPS00,GPS01} of the spin fluctuations by bosonic spinons. The bosonic spinon theory is valid for large $S$, and so leads to a large $q_{EA}$ (see Appendix~\ref{app:bosons}); it also requires an additional assumption of marginal stability of a replica symmetry breaking solution to obtain the gapless spectrum. 
Our analysis uses fermionic spinons, does not require any additional marginal stability criteria, and is applicable for 
small $q_{EA}$. It is reassuring that the same qualitative behavior is obtained by fermionic and bosonic spinons. Thus we have a `duality' between fermionic and bosonic spinons present not only in the gapless, fractionalized, spin liquid regime \cite{SY92}, but also in the crossover to the confining spin glass state. We note that boson-fermion dualities have seen much discussion in the context of disorder-free gapless spin liquids on the square lattice \cite{Wang2017,Thomson:2017ros}. 

We will begin in Section~\ref{sec:largeN} by formulating the path integral of the random SU($M$) magnet for large $N$ but general $M$. This will be a $G$-$\Sigma$-$Q$ theory, 
involving a path integral over the fermionic spinon Green's function and self energy, $G,\Sigma$, and the spin auto-correlation $Q$. We will present the $M \rightarrow \infty$ limit of this theory in Section~\ref{sec:largeM}, which yields the spin liquid state of Ref.~\cite{SY92}. Spin glass order is absent at $M=\infty$, but is present at any finite $M$ because of a logarithm-squared divergence of the spin glass susceptibility \cite{GPS01}. We will describe such finite $M$ effects in Section~\ref{sec:1M}, and present the structure of the effective action for $q_{EA}$ and the spin autocorrelation function in powers of $1/M$. Section~\ref{sec:spectrum} combines our results to obtain the feedback of the spin glass order on the dynamic spin spectrum. The low temperature complexity of the quantum spin glass state is discussed in Section~\ref{sec:complexity}.

\section{Large $N$ action}
\label{sec:largeN}

All our analysis will be carried out in the $N \rightarrow \infty$ limit of a model with SU($M$) symmetry. We will keep $M$ arbitrary in the present section.
We consider the SU($M$) spin model
\beq
H = \frac{1}{\sqrt{NM}} \sum_{i < j = 1}^N \sum_{\alpha,\beta=1}^M J_{ij} \mathcal{S}_\beta^\alpha (i) \mathcal{S}_\alpha^\beta (j) \label{SUMHam}
\eeq
where $\mathcal{S}_\beta^\alpha (i) = [\mathcal{S}_\alpha^\beta (i)]^\dagger$ are generators of SU($M$) on each site $i$, with $\alpha, \beta = 1 \ldots M$. Each site contains states corresponding to the {\it antisymmetric\/} product of $k M$ (integer) fundamentals, and these are realized by fermionic spinons with
\beq
\mathcal{S}_{\beta}^\alpha (i) = f_\beta^\dagger (i) f^\alpha (i) - k \delta^{\alpha}_{\beta} \,,\quad \sum_\alpha  f_\alpha^\dagger (i) f^\alpha (i) = k M \label{SkM}
\eeq
with fermions $f^\alpha (i)$ on each site $i$; the model with bosons on each site realized the {\it symmetric\/} product of fundamentals, and is briefly discussed in Appendix~\ref{app:bosons}. Note that (\ref{SkM}) implies that the spinons carry a U(1) gauge charge (see Appendix~\ref{app:U1}), unlike the fermions of the SYK model. We have made the spin operators traceless, and will restrict ourselves to the particle-hole symmetric case $k=1/2$. The Hamiltonian in (\ref{SUMHam}) reduces to the $S=1/2$ case of the SU(2) Hamiltonian in (\ref{SU2Ham}) for $M=2$ and $k=1/2$ (apart from an overall factor of $1/\sqrt{2}$).

We introduce replicas $a=1 \ldots n$, and average over $J_{ij}$ to obtain the averaged, replicated partition function
\bea
\overline{\mathcal{Z}^n} &=& \int \mathcal{D} f_a^\alpha (i, \tau)  \mathcal{D} \lambda_{a} (i, \tau) \exp \left[ - {S}_B - {S}_J \right] \nonumber \\
{S}_B &=& \sum_i \int d \tau \left[ f_{a\alpha}^\dagger (i) \partial_\tau f_{a}^\alpha (i) + i \lambda_a (i) \left( f_{a\alpha}^\dagger (i) f_a^\alpha (i) - k M \right) \right] \nonumber \\
{S}_J &=& - \frac{J^2}{4NM} \int d \tau d \tau' \left[ \sum_i \mathcal{S}_{a \beta}^\alpha (i, \tau) \mathcal{S}_{b \delta}^\gamma (i,\tau') \right]
\left[ \sum_j \mathcal{S}_{a \alpha}^\beta (j, \tau) \mathcal{S}_{b \gamma}^\delta (j,\tau') \right]
\eea
We can now decouple $\mathcal{S}_J$ with a Hubbard-Stratonovich field $Q_{ab,\beta\delta}^{\alpha\gamma} (\tau, \tau')$ and take the large $N$ limit.
Then the problem reduced to finding saddle points of the single site action
\beq
\frac{\mathcal{S}[Q]}{N} =  \frac{J^2}{4 M} \int d \tau d \tau' |Q_{ab,\beta\delta}^{\alpha\gamma} (\tau, \tau')|^2 - \ln \mathcal{Z}_f [Q] \label{SQ}
\eeq
where $\mathcal{Z}_f [Q]$ is the single site partition function
\bea
\mathcal{Z}_f [Q] &=& \int \mathcal{D} f_a^\alpha (\tau)  \mathcal{D} \lambda_{a} (\tau) \exp \left[ - {S}_B - {S}_f \right] \label{ZQ1} \\
{S}_B &=& \int d \tau \left[ f_{a\alpha}^\dagger \partial_\tau f_{a}^\alpha  + i \lambda_a  \left( f_{a\alpha}^\dagger  f_a^\alpha - k M \right) \right] 
\label{ZQ2} \\
{S}_f &=& - \frac{J^2}{2M} \int d \tau d \tau' Q_{ab,\beta\delta}^{\alpha\gamma} (\tau, \tau') \left[ f_{a \alpha}^\dagger (\tau) f_a^\beta (\tau) - k \delta_{\alpha}^\beta \right] \left[ f_{b \gamma}^\dagger (\tau') f_b^\delta (\tau') - k \delta_{\gamma}^{\delta} \right] \label{ZQ3}
\eea
Note that now there is no remaining path integral over $Q$. We simply have to find the saddle points of the action $\mathcal{S}[Q]$ in (\ref{SQ}). 

Let us assume that the saddle point does not break spin rotation symmetry: this is true in both the spin glass, and spin liquid phases. So we make the ansatz \cite{SY92}
\beq
Q_{ab,\beta\delta}^{\alpha\gamma} (\tau, \tau') = \delta^{\alpha}_\delta \delta^\gamma_\beta \, Q_{ab} (\tau- \tau') \label{Qinv}
\eeq	
where $Q_{ab} (\tau)$ is a real function. Also, because there is no path integral over $Q$, we can also assume from now on that $Q_{ab} (\tau)$ is independent of $\tau$ for $a \neq b$ \cite{RSY95}.
Then (\ref{SQ}) is replaced by
\beq
\frac{\mathcal{S}[Q]}{N} =  \frac{J^2 M}{4} \int d \tau d \tau' [Q_{ab}(\tau- \tau')]^2 - \ln \mathcal{Z}_f [Q] \label{SQ1}
\eeq
while (\ref{ZQ3}) is replaced by
\beq
{S}_f = - \frac{J^2}{2M} \int d \tau d \tau' Q_{ab} (\tau-\tau') \left[ f_{a \alpha}^\dagger (\tau) f_a^\beta (\tau) f_{b \beta}^\dagger (\tau') f_b^\alpha (\tau') 
 - k^2 M  \right]\label{ZQ4}
\eeq
Finally, we express $\mathcal{Z}_f[Q]$ as a $G$-$\Sigma$ theory \cite{Maldacena:2016hyu,Kitaev:2017awl}. We define the spinon Green's function
\beq
G_{ab} (\tau, \tau') = - \frac{1}{M} \sum_\alpha f_a^{\alpha} (\tau) f_{b \alpha}^\dagger (\tau') \,.
\eeq
Then we can write
\beq
\mathcal{Z}_{f} [Q] = \exp \left(  - \frac{k^2 J^2}{2}  \int d \tau d \tau' \sum_{a,b} Q_{ab} (\tau-\tau') \right) \int \mathcal{D} G_{ab} (\tau, \tau') \mathcal{D}\Sigma_{ab} (\tau, \tau')  \mathcal{D} \lambda_{a} (\tau) \, \exp \left[ - M I[Q] \right] \label{ZQ5}
\eeq
where the action $I[Q]$ is
\bea
I[Q] &=& -  \ln \mbox{det} 
 \biggl[ - \delta' (\tau - \tau') \delta_{ab} - i \lambda_a (\tau) \delta(\tau - \tau') \delta_{ab}  - \Sigma_{ab} (\tau, \tau') \biggr] - i k \int d \tau \lambda_a (\tau)   \nonumber \\
&~& + \int d \tau d \tau' \left[ - \Sigma_{ab} (\tau, \tau') G_{ba} (\tau', \tau) + \frac{J^2}{2} Q_{ab} (\tau - \tau') G_{ab} (\tau, \tau') G_{ba} (\tau', \tau) \right] \,.  \label{ZfQ}
\eea

We note that (\ref{ZQ5}) and (\ref{ZfQ}) constitute an exact formulation of the theory for all $M$. Our remaining task is to evaluate the path integral over $G_{ab} (\tau, \tau')$, $\Sigma_{ab} (\tau, \tau')$, and $\lambda_a (\tau)$ in (\ref{ZQ5}), and then determine the saddle-point solutions for $Q_{ab} (\tau)$ in (\ref{SQ1}). 
The saddle point equations for $Q$ from (\ref{SQ1}), (\ref{ZQ4}), and (\ref{ZfQ}) are
\bea
Q_{ab} (\tau - \tau') &=& \frac{1}{M^2} \left\langle f_{a \alpha}^\dagger (\tau) f_a^\beta (\tau) f_{b \beta}^\dagger (\tau') f_b^\alpha (\tau') \right\rangle_{\mathcal{Z}_f [Q]} - \frac{k^2}{M} \nonumber \\
&=& - \Bigl\langle G_{ab} (\tau, \tau') G_{ba} (\tau', \tau) \Bigr\rangle_{\mathcal{Z}_f [Q]} - \frac{k^2}{M}\,, \label{ZQ6}
\label{Qf}
\eea
but we will find it more convenient to obtain them directly from the functional form of $\mathcal{S}[Q]$. 

From the resulting $Q_{ab} (\tau)$, we obtain two different characterizations of the spin glass order \cite{YSR93,RSY95,GPS00,GPS01,Cugliandolo00,Anous:2021eqj}. At $T=0$, we can examine the long-time limit of the replica diagonal $Q$
\beq
\overline{q} = \lim_{n \rightarrow 0} \frac{1}{n} \sum_a Q_{aa} (\tau \rightarrow \infty)\,, \quad T=0\,,
\label{eq:qbarDef}
\eeq
and $\overline{q}$ is one measure of the spin-glass order. Alternatively, we can examine the off-diagonal components, which are necessarily time-independent
\beq
q_{ab} = Q_{ab} (\tau) \quad, \quad a \neq b\,.
\label{eq:qabDef}
\eeq
In the $n \rightarrow 0$ limit, it is conventional to describe the ultra-metric structure of $q_{ab}$ by the Parisi function $q(x)$, $0 \leq x \leq 1$, and the Edwards-Anderson spin glass order parameter is $q_{EA} = q(1)$. Consistency between the two different characterizations requires that $\overline{q} = q_{EA}$, and this is an important feature of earlier studies of quantum spin glasses \cite{RSY95}. 

These definitions also allow us to place a bound on spin-glass order. The state with maximum order has the spin frozen in a state in which the fermions occupy the states with, say, $\alpha = 1 \ldots k M$, while the other values of $\alpha$ are empty. Evaluating (\ref{Qf}) on such a state, we obtain
\beq
q_{EA} \leq \frac{k(1-k)}{M}\,. \label{maxq}
\eeq
Note that (\ref{maxq}) vanishes as $M \rightarrow \infty$, and $q_{EA}$ is at most $\mathcal{O}(1/M)$ in the large $M$ limit; this is consistent with our results in Sections~\ref{sec:largeM} and \ref{sec:1M}. In Appendix~\ref{app:bosons} we review the bosonic spinon case of (\ref{SUMHam}), and find there that $q_{EA}$ can be $\mathcal{O}(M^0)$ in that large $M$ limit. We also note that for SU(2), the definition of the spin glass order from (\ref{ZQ6}) is $q_{EA} = \langle {\bm S}_i \rangle \cdot \langle {\bm S}_i \rangle/2$, and this is a factor of 2 smaller than the usual definition; so the bound in (\ref{maxq}) is $q_{EA} \leq 1/8$.

We would now like to evaluate $\ln \mathcal{Z}_f [Q]$ for general $Q_{ab} (\tau)$, with $Q_{ab}$ independent of $\tau$ for $a \neq b$.
We first do this at $M = \infty$ in Section~\ref{sec:largeM}, and then examine $1/M$ corrections in Section~\ref{sec:1M}.

\section{Large $M$ limit}
\label{sec:largeM}

Assuming a general $Q_{ab} (\tau)$, the large $M$ limit of the path-integral in (\ref{ZQ5}) leads to the following saddle-point equations for the fermion Green's function and self-energy
\bea
\Sigma_{ab} ( \tau) &=& J^2 Q_{ab} (\tau) G_{ab} (\tau) \nonumber \\
G_{ab} (i \omega) &=& \left[ i \omega \delta_{ab} - \Sigma_{ab} (i \omega) \right]^{-1} \label{e1}
\eea
where $\lambda_a = 0$ at the $k=1/2$ saddle-point because of particle-hole symmetry. However, we must keep in mind that there cannot be any off-diagonal components of the fermion Green's function at the saddle-point, because it is not possible for fermions to condense. So we write
\beq
G_{ab} (\tau, \tau') = G_Q (\tau - \tau') \delta_{ab} \quad,\quad M=\infty\,, \label{e1c}
\eeq
and similarly for $\Sigma_{ab}$. From the large $N$ saddle-point equation for $Q_{ab}$ in (\ref{ZQ6}), we see that $Q_{ab}$ must also be replica diagonal,
\beq
Q_{ab} (\tau) = Q (\tau) \delta_{ab} \quad,\quad M=\infty\,,
\eeq
and so there is no spin glass order at $M=\infty$ \cite{SY92}. 
The large $M$ saddle point equations (\ref{e1}) therefore reduce to
\bea
\Sigma_Q ( \tau) &=& J^2 Q(\tau) G_Q (\tau) \nonumber \\
G_Q (i \omega) &=& \left[ i \omega - \Sigma_Q (i \omega) \right]^{-1} \,. \label{e1a}
\eea
These equations hold for general $Q(\tau)$, and we have emphasized this by the subscript $Q$ on $G$ and $\Sigma$. Upon including the large $N$ saddle point equation for $Q$ in (\ref{ZQ6}), we obtain
\beq
Q(\tau) = - G_Q (\tau) G_Q (-\tau)\,,\quad M=\infty\,. \label{e1b}
\eeq
The combination of (\ref{e1a}) and (\ref{e1b}) yields precisely the large $N$ equations of the fermion of the complex SYK model \cite{SY92}. In the following sections, we include corrections from the replica off-diagonal and two-time fluctuations of $G_{ab} (\tau, \tau')$ and $\Sigma_{ab} (\tau, \tau')$, and these will modify (\ref{e1b}), but we will continue to use (\ref{e1a}).

For completeness, we also present the expressions for the path integral in
 (\ref{ZfQ}):
\bea
- \frac{\ln \mathcal{Z}_f [Q]}{Mn} &=& \frac{I[Q]}{n} + \frac{k^2 J^2}{2Mn}  \int d \tau d \tau' \sum_{a,b} Q_{ab} (\tau-\tau') \nonumber \\
\frac{I[Q]}{n} &=& -  \ln \mbox{det} 
 \biggl[ - \delta' (\tau - \tau')   - \Sigma_Q (\tau - \tau') \biggr]   \label{ZfQa} \\
&+& \int d \tau d \tau' \left[ - \Sigma_Q (\tau - \tau') G_Q (\tau' - \tau) + \frac{J^2}{2} Q(\tau - \tau') G_Q (\tau- \tau') G_Q (\tau'- \tau) \right]\,. \nonumber
\eea

\section{$1/M$ expansion}
\label{sec:1M}

This section will describe $1/M$ corrections to $\ln \mathcal{Z}_f [ Q]$ in (\ref{SQ1}). We will see below that these corrections are characterized by a divergent spin glass susceptibility, and so spin glass order is present for any finite $M$ \cite{GPS01}.

To evaluate these finite $M$ fluctuations, 
we extend (\ref{e1c}) for the fermion Green's function and self-energy, and for the constraint Lagrange multiplier by
\bea
G_{ab} (\tau, \tau') &=&  G_Q (\tau-\tau') \delta_{ab}  + \delta G_{ab} (\tau, \tau') \nonumber \\
 \Sigma_{ab} (\tau, \tau') &=&  \Sigma_Q (\tau-\tau') \delta_{ab} + \delta \Sigma_{ab} (\tau, \tau') - i \delta \lambda_a (\tau) \delta(\tau - \tau') \delta_{ab} \nonumber \\
 \lambda_a (\tau) &=& \overline{\lambda}_a + \delta \lambda_a (\tau) \,, \label{e5}
\eea
where $\overline{\lambda}_a = 0$ at the $M=\infty$ saddle point for the particle-hole symmetric case $k=1/2$. We can use the gauge invariance discussed in Appendix~\ref{app:U1} to choose a gauge in which $\delta \lambda_a (\tau)$ is $\tau$ independent. Then the time-independent value of $\delta \lambda_a$ can be absorbed into $\overline{\lambda}_a$, and evaluating the path integral over $\delta \lambda_a (\tau)$ to relative order $1/M^2$ reduces to computing the shift in the saddle-point value of $\overline{\lambda}_a$ to order $1/M$ \cite{ReadNewns,CSY}. This shift in the value of $\overline{\lambda}_a$ has to be included in $G_Q$.
Also, while the expectation values of $G_{ab} (\tau, \tau')$, $\Sigma_{ab} (\tau, \tau')$ must depend only upon $\tau-\tau'$ and have to be replica diagonal, the fluctuations $\delta G_{ab} (\tau, \tau')$, $\delta \Sigma_{ab} (\tau, \tau')$ of both replica diagonal and replica off-diagonal components must include full dependence on both $\tau$ and $\tau'$.

\subsection{Determinant of quadratic fluctuations}

We begin with the first $1/M$ corrections, which are associated with quadratic fluctuations of $\delta G_{ab} (\tau, \tau')$, $\delta \Sigma_{ab} (\tau, \tau')$. Expanding the action \eqref{ZfQ} to second order in fluctuations around the large $M$ saddle point, we find the quadratic action
\begin{equation}
\label{eq:fluctInt}
\begin{split}
  &I_{(2)}[Q] 
  = \frac{1}{2} \int d\tau_1 \cdots d\tau_4 \sum_{a,b,c,d}\delta X^T_{ab}(\tau_1,\tau_2) \cdot A_{ab;cd}(\tau_1,\tau_2;\tau_3,\tau_4) \cdot \delta X_{cd} (\tau_3,\tau_4) 
  \equiv \frac{1}{2} \, \delta {\bf X}^T \cdot {\bf A} \cdot \delta {\bf X}
  \end{split}
 \end{equation}
 where matrix multiplication involves the following structures:
 {\small
 \begin{equation}
 \label{eq:flucts}
 \begin{split}
 \delta {\bf X} \equiv \delta X_{ab}(\tau_1,\tau_2) &\equiv \begin{pmatrix} \delta G_{ab}(\tau_1,\tau_2)  \,,\\
 \delta \Sigma_{ab}(\tau_1,\tau_2)  
 \end{pmatrix}
\,, \\
{\bf A} \equiv A_{ab;cd}(\tau_1,\tau_2;\tau_3,\tau_4) &\equiv       
 \begin{pmatrix} J^2  \, Q_{ab}(\tau_1,\tau_2) \delta(\tau_{14}) \delta(\tau_{32})
 & - \delta(\tau_{14})\delta(\tau_{32})  \\
  -\delta(\tau_{14})\delta(\tau_{32})
  &  G_Q(\tau_{14}) G_Q(\tau_{32})
 \end{pmatrix} \delta_{ad}\delta_{bc} \,,
 \end{split}
\end{equation} 
and the dot product is defined as indicated in terms of integration over pairs of time arguments and summation over pairs of replica indices.
\begin{figure}
    \centering
    \includegraphics[width=0.42\textwidth]{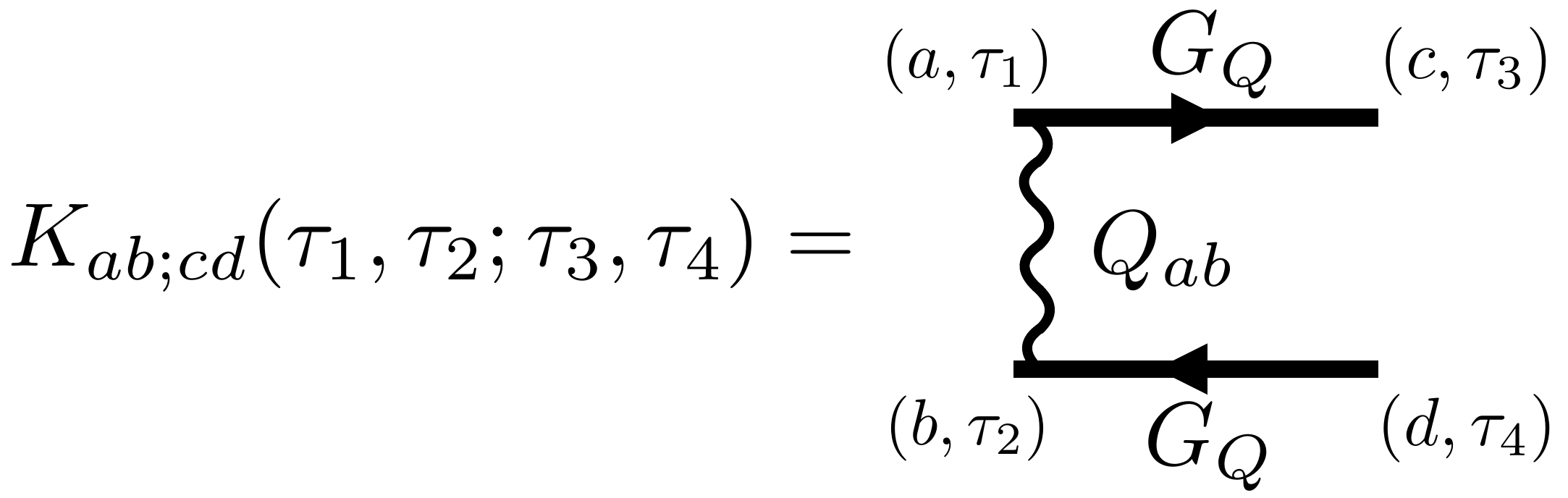}
    \caption{Pictorial representation of the `ladder' kernel featuring in the fluctuation determinant.}
    \label{fig:diagrams0}
\end{figure}

Having reduced the problem to a Gaussian integral, we are now in a position to evaluate the contribution to the free energy $ {\cal F} = - (M\beta n)^{-1} \ln  {\cal Z}_f[Q]$ originating from quadratic fluctuation determinants. We denote this contribution as
\begin{equation}
\label{eq:lndet}
\beta n \, {\cal F}_{sg}  =  \frac{k^2J^2}{2} \int d\tau d\tau' \sum_{a,b} Q_{ab}(\tau,\tau') + \frac{1}{2} \,\ln \det {\bf A} 
+ {\cal O}(M^{-1})\,.
\end{equation}
Let us focus on the fluctuation determinant of ${\bf A}$, which can be expanded in terms of a ladder kernel:
\begin{equation}
\frac{1}{2}\,\ln \det {\bf A} 
 =   \frac{1}{2}\,  \text{Tr} \,\ln({\bf K}-{\bf 1}) 
 =  - \frac{1}{2} \, \text{Tr} \, {\bf K} - \frac{1}{4} \, \text{Tr}\, {\bf K}^2 - \frac{1}{6} \, \text{Tr}\, {\bf K}^3 + \ldots
\end{equation}
where the ladder kernel and the identity operator are defined as
\begin{equation}
\begin{split}
 {\bf K} &\equiv K_{ab;cd}(\tau_1,\tau_2;\tau_3,\tau_4) \equiv 
J^2\, Q_{ab}(\tau_1,\tau_2) G_Q(\tau_1-\tau_3)G_Q(\tau_4-\tau_2)\,\delta_{ac}\delta_{bd}\,, \\
{\bf 1} &\equiv \delta(\tau_1-\tau_3)\delta(\tau_4-\tau_2) \, \delta_{ac} \delta_{bd} \,.
\end{split}
\end{equation}
In Fig.~\ref{fig:diagrams0} we introduce a diagrammatic notation for the kernel.\footnote{ The diagrams focus on the structure of replica indices. To recover the fermionic description one uses a double line notation where the wiggly line fattens into two lines carrying SU($M$) indices.}
Fig.~\ref{fig:diagrams1} further illustrates the above contributions to the free energy diagrammatically. One can check that the first diagram, $\text{Tr}\, {\bf K}$, is cancelled by the first term in \eqref{eq:lndet}.\footnote{ In the computation of $\text{Tr}\, {\bf K}$ we use the following point splitting prescription to be consistent with the fermionic description: 
\begin{equation}
    \text{Tr}\,{\bf K} = \lim_{\varepsilon \rightarrow 0 } \int d\tau_1 d\tau_2  \, \sum_{a,b} {\bf K}_{ab;ab}(\tau_1,\tau_2;\tau_1+\varepsilon,\tau_2-\varepsilon)\,.
\end{equation}
}
The traces of higher powers of the ladder kernel yield:
\begin{equation}
\begin{split}
 -\frac{1}{4} \text{Tr}\, {\bf K}^2 
   &= -\frac{J^4}{4} \int d\tau_1 \cdots d\tau_4 \sum_{a,b} Q_{ab}(\tau_1,\tau_2) Q_{ab}(\tau_3,\tau_4)  \, R_Q^{(2)}(\tau_{13}) R_Q^{(2)}(\tau_{24})\,, \\
 -\frac{1}{6} \text{Tr}\, {\bf K}^3 
   &= \frac{J^6}{6} \int d\tau_1 \cdots d\tau_6 \sum_{a,b} Q_{ab}(\tau_1,\tau_2) Q_{ab}(\tau_3,\tau_4)Q_{ab}(\tau_5,\tau_6)  \, R_Q^{(3)}(\tau_{13},\tau_{35}) R_Q^{(3)}(\tau_{24},\tau_{46})\,,
\end{split}
\end{equation}
and so on, where the time splitting functions are given by the spinon loops
\begin{equation}
\begin{split}
 R_Q^{(2)}(\tau) &\equiv G_Q(\tau) G_Q(-\tau) \,,\\
 R_Q^{(3)}(\tau,\tau') &\equiv G_Q(\tau) G_Q(\tau') G_Q(-\tau-\tau') \,,\quad \ldots
 \end{split}
\end{equation} 
\begin{figure}
    \centering
    \includegraphics[width=0.8\textwidth]{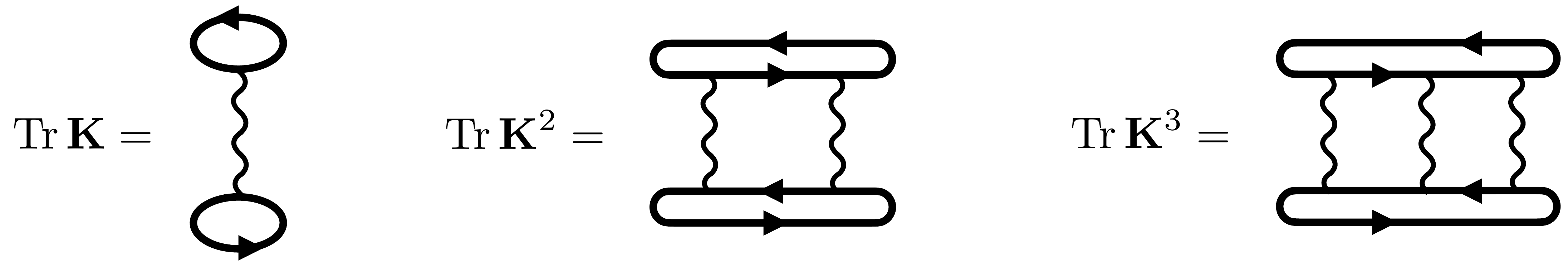}
    \caption{Some diagrams contributing to $-\ln {\cal Z}_f[Q]$ at the first subleading order, i.e., ${\cal O}(M^0)$.}
    \label{fig:diagrams1}
\end{figure}

Consider now the ansatz for $Q_{ab}$, which we described in \eqref{eq:qbarDef} and \eqref{eq:qabDef}. Without loss of generality, we parameterize this ansatz as follows:
\begin{equation}
 Q_{ab}(\tau,\tau') = \left[ Q(\tau-\tau') + \overline{q} \right]\, \delta_{ab} + q_{ab} \,, \qquad q_{aa} =0 \,. \label{Qq}
\end{equation}
We then find a simple expression for the contribution of ${\cal Z}_f[Q]$ to the free energy per spin. In particular, the subleading terms described above yield a contribution to the free energy, which we denote as 
\begin{equation}
\label{eq:Fsg}
\begin{split}
 {\cal F}_{sg} \equiv  - \frac{\ln {\cal Z}_f[Q]}{\beta n} 
 &= -c_0 - c_{1}   \overline{q} 
 -
 c_{2} \overline{q}^2 - d_{2}  \beta \left( \overline{q}^2 +\frac{1}{n} \sum_{a\neq b} q_{ab}^2 \right) -
 c_{3} \overline{q}^3 -
 c_{4} \overline{q}^4 - d_{4} \, \beta \left( \overline{q}^4 + \frac{1}{n}\sum_{a\neq b} q_{ab}^4 \right) + \ldots 
\end{split}
\end{equation}
where we organize the expression as an expansion in powers of $\overline{q}$ and $q_{ab}$. The coefficients are given by
\begin{equation}
        d_{2} = \frac{J^4}{4} \, R_Q^{(2)}(\omega=0)^2 \,, \qquad
        d_{4} = \frac{J^8}{8} \, R_Q^{(4)}(\omega_1=0,\,\omega_2=0,\,\omega_3=0)^2 \,, \qquad \ldots 
 \end{equation}
 and
\begin{equation}
    \begin{split}
        c_0 &=  
        \frac{J^4}{4\beta} \sum_\omega R_Q^{(2)}(\omega)^2 Q(\omega)^2 + \frac{J^6}{6\beta^2} \sum_{\omega,\omega'} R_Q^{(3)}(\omega,\omega')^2 Q(\omega) Q(\omega-\omega') Q(\omega') \\
        &\quad + \frac{J^8}{8\beta^3} \sum_{\omega,\omega',\omega''} R_Q^{(4)}(\omega,\omega',\omega'')^2 Q(\omega) Q(\omega-\omega')Q(\omega'-\omega'') Q(\omega'') 
        + \ldots \, \\
        c_{1}  &= \frac{J^4}{2}\, R_Q^{(2)}(0)^2\, Q(0) + \frac{J^6}{2\beta} \, \sum_{\omega}  R_Q^{(3)} (\omega,0)^2 Q(\omega)^2 +\frac{J^8}{2\beta^2} \sum_{\omega,\omega'} R_Q^{(4)}(\omega,\omega',0)^2Q(\omega)Q(\omega-\omega')Q(\omega')+ \ldots \\
        c_{2} &= 
        \frac{J^8}{4\beta}\, \sum_{\omega}  \left[ 2 R_Q^{(4)} (\omega,0,0)^2 + R_Q^{(4)} (\omega,\omega,0)^2 \right] Q(\omega)^2 + \ldots \,,\\
        c_3 &= \frac{J^8}{2} \, R_Q^{(4)}(0,0,0)^2 Q(0)+ \ldots
    \end{split}
\end{equation}
where all $Q$ and $R_Q^{(n)}$ in the above equations are frequency space expressions. We also used their symmetry properties,
\begin{equation}
 R_Q^{(n)}(-\omega_1,\ldots,-\omega_{n-1})=  (-1)^n \, R_Q^{(n)}(\omega_1,\ldots,\omega_{n-1}) \,,
\end{equation}
to simplify some expressions and to conclude that coefficients such as $d_3 = d_5 = \ldots = 0$ (for the particle-hole symmetric case $k=1/2$). Note that we have reinstated explicit $\beta$-dependence in the above formulas in order to make manifest that only the terms multiplying $d_k$ are linearly proportional to $\beta$ in the low temperature limit, $\beta \rightarrow \infty$. In general, we find that the coefficients of these linearly divergent terms are always negative and given by
\begin{equation}
 d_{2k} = \frac{J^{4k}}{4k} \left( \frac{1}{\beta}\sum_\omega G_Q(\omega)^{2k} \right)^2 \qquad\quad (k = 1,2,\ldots). \label{d2kval}
\end{equation}
When evaluated on the spin liquid Green's function $G_Q (\omega) \sim 1/\sqrt{\omega}$, we find a further divergence in the values of $d_{2k}$
in (\ref{d2kval}): $d_{2k} \sim \beta^{2k-2}$. However, this divergence is cutoff when we compute $d_{2k}$ using the self-consistent results for $G_Q (\omega) $ to be computed in Section~\ref{sec:spectrum}: the cutoff frequency scale is $\omega_\star$ in (\ref{omegastar}), and hence $d_{2k} \sim q_{EA}^{2-2k}$. The net contribution of all the $d_{2k}$ terms in (\ref{eq:Fsg}) to the free energy is therefore of order $\beta q_{EA}^2$.
For $k=1$, there is an additional logarithm of $\beta$ (or $\omega_\ast$), as noted below in (\ref{chiloc1}).

Higher orders in the $1/M$ expansion can be computed in a similar fashion. In short, these are characterized by more complicated diagrams build from the kernel ${\bf K}$. We elaborate on this in Appendix~\ref{app:higherOrders}.

\subsection{Free energy}
\label{sec:freeen}

In order for the theory to be consistent we will need to ensure that physical quantities such as the free energy are finite as $\beta \rightarrow \infty$. As we discuss next, this follows indeed from the equations of motion for the spin glass parameters $\overline{q}$ and $q_{ab}$.

The free energy including the corrections to first subleading in the $1/M$ expansion reads as follows:
\begin{equation}
\beta n \,{\cal F} \equiv\frac{{\cal S}[Q]}{NM} = \frac{\beta n J^2}{4}  \left[\frac{1}{\beta} \sum_\omega Q(\omega)^2 + 2\overline{q} \, Q(\omega=0) + \beta \left( \overline{q}^2 + \frac{1}{n} \sum_{a\neq b} q_{ab}^2 \right)\right]  + I[Q] + \frac{\beta n}{M} \, {\cal F}_{sg} + {\cal O}(M^{-2}) \,,
\end{equation}
where the leading terms were given in \eqref{SQ1} and \eqref{ZfQa}, while the $\mathcal{F}_{sg}$ term was computed in \eqref{eq:Fsg}. Of particular importance is the term quadratic in the spin glass order parameter
\beq
\frac{{\cal S}[Q]}{NM} = \frac{\beta^2 n J^2}{4}\left( \overline{q}^2 + \frac{1}{n} \sum_{a\neq b} q_{ab}^2 \right) \left[ 1 - \frac{J^2}{M} \chi_{\rm loc}^2 \right] + \ldots\,, \label{chisg}
\eeq
where $\chi_{\rm loc} =  - R_Q^{(2)}(\omega=0)$ is the local spin susceptibility. The term in square brackets in (\ref{chisg}) is precisely that appearing in the denominator of the spin glass susceptibility \cite{GPS01}. In the SYK spin liquid state \cite{SY92} (this is evident from the Hilbert transform of (\ref{chisgn})), 
\beq
\chi_{\rm loc} = \int_0^\beta Q(\tau) d \tau = \frac{1}{ J \sqrt{\pi}} \ln (\beta J) \,, \label{chiloc1}
\eeq
 and so the term in square brackets becomes negative at low enough temperatures provided $M$ is finite. Once this term is negative, spin glass order will appear, and we obtain an estimate 
 \beq
 T_c \sim J \exp \left(- \sqrt{M \pi} \right) 
 \label{Tc}
 \eeq
 for the critical temperature \cite{GPS01}. For temperatures below $T_c$, $\chi_{\rm loc}$ is finite at $T=0$ in the presence of spin glass order, as we will see in Section~\ref{sec:spectrum}.

The simplest ansatz for evaluating the free energy assumes a replica symmetric off-diagonal spin glass order of the form $q_{a\neq b} = q_{EA}$. In this case, we employ the following simplification as $n\rightarrow 0$:
\begin{equation}
\frac{1}{n} \sum_{a\neq b} q_{ab}^\ell = (n-1) q_{EA}^\ell \quad \longrightarrow \quad -q_{EA}^\ell \,.
\end{equation}
Extremization of ${\cal F}$ with respect to $q_{EA}$ then yields the following equation of motion:
\begin{equation}
    \frac{J^2}{2} \, q_{EA} = \frac{1}{M} \left[  2d_2 \, q_{EA} + 4 d_4 \, q_{EA}^3 + \ldots \right] + {\cal O}(M^{-2})\,.
\end{equation}
Similarly, extremization with respect to $\overline{q}$ gives:
\begin{equation}
    \frac{J^2}{2} \, \left[ \overline{q} +\frac{1}{\beta} \, Q(0) \right] =  \frac{1}{M} \left[   2d_2 \, \overline{q} + 4 d_4 \, \overline{q}^3 + \ldots + \frac{1}{\beta} \left( c_1 + 2 c_2\,\overline{q} + 3c_3\,\overline{q}^2 + 4 c_4\,\overline{q}^3    + \ldots\right) \right] + {\cal O}(M^{-2})\,.
\end{equation}
Evidently, these equations imply
\begin{equation}
\label{eq:qbarqEA}
    \overline{q} = q_{EA} + {\cal O}(\beta^{-1}) \,.
\end{equation}
Evaluated on this solution, the free energy is indeed finite as $\beta \rightarrow \infty$ since all dangerous terms are of the following form as $n\rightarrow 0$:
\begin{equation}
   {\cal F} =  \beta \left\{ \frac{ J^2}{4} (\overline{q}^2-q_{EA}^2 ) 
    - \frac{1}{M} \sum_{k\geq 1} d_{2k} \left( \overline{q}^{2k}-q_{EA}^{2k} \right)
    + {\cal O}(M^{-2})\right\}+ {\cal O}(\beta^0)  = {\cal O}(\beta^0) \,,
\end{equation}
where in the last step we used the relation \eqref{eq:qbarqEA}. In Appendix~\ref{app:higherOrders} we compute some examples of contributions at higher orders in the $1/M$ expansion, and show that these also have a finite limit as $\beta \rightarrow \infty$.

A notable feature of this analysis is that the free energy is finite in the $\beta \rightarrow \infty$ limit, even though there are many individual terms that diverge in this limit. There is a delicate cancellation of the divergent terms between the replica diagonal and off-diagonal contributions in the $n \rightarrow 0$ limit \cite{RSY95}.
This cancellation was overlooked in an early work on the random quantum magnet \cite{BrayMoore}: they only included the replica diagonal terms, which in fact diverge as $\beta \rightarrow \infty$, and so their energy estimates are not meaningful. Such divergent contributions to the free energy are also present in various EDMFT theories of strongly correlated phases \cite{Werner10,Akerlund:2013fsa,Akerlund:2014mea,Si1,SiIsing2,SiIsing3}, and we believe that the energy estimates in such theories are not reliable in the phase with long-range order at very low temperatures.

\section{Spectrum of the spin glass state}
\label{sec:spectrum}

We have seen in Section~\ref{sec:1M} that the order parameter characterizing the spin glass ground state, $q_{ab}$, is determined entirely by corrections to the leading large $M$ saddle point. Moreover, as $\beta \rightarrow \infty$, the long time limit of the spin autocorrelation function, $\overline{q}$, equals the Edwards-Anderson order parameter $q_{EA}$ (which is in turn determined from $q_{ab}$). In this section, we will address the feedback of the onset of spin glass order on the spinon Green's function and the dynamic spin susceptibility.

In Section~\ref{sec:largeM}, we determined the large $M$ equations, (\ref{e1a}), obeyed by the fermion Green's function for a general spin autocorrelation function $Q (\tau)$. In the spin glass phase, we mapped $Q (\tau) \rightarrow Q(\tau) + \overline{q}$ in (\ref{Qq}) to allow for a non-zero long time limit. The computations of Section~\ref{sec:1M}, will lead to corrections to $Q(\tau)$ at order $1/M$, along with allowing for a non-zero $\overline{q}$. In our analysis here, we will ignore the $1/M$ corrections to $Q(\tau)$, as they have a structure similar to that obtained in the $M = \infty$ theory. However, we will keep the non-zero value of $\overline{q} = q_{EA}$ because it has a singular effect on the low frequency fermion spectrum, as we will now show.

The upshot of this discussion is that we can determine the fermion Green's function by solving (\ref{e1a}), while (\ref{e1b}) is modified to 
\beq
Q(\tau) = - G_Q (\tau) G_Q (-\tau) + q_{EA}\,. \label{e1bM}
\eeq
Remarkably, the equations (\ref{e1a}) and (\ref{e1bM}) have been solved previously \cite{PG98, Balents2017}, in different contexts. Ref.~\cite{PG98} considered a random $t$-$J$ model in a particular large $M$ limit,  with r.m.s. exchange $J$, and r.m.s. hopping $t$. Ref.~\cite{Balents2017} considered a SYK model with a random 4-fermion interaction term with r.m.s. strength $J$, and a random 2-fermion hopping term $t$. The equations of the latter model map onto (\ref{e1a}) and (\ref{e1bM}) with $t = J \sqrt{q_{EA}}$.
Their main result was that there was a crossover from SYK non-Fermi liquid behavior to Fermi liquid behavior at a coherence energy scale $\sim t^2/J$ \cite{PG98, Balents2017} which equals $J q_{EA}$. From this, we can obtain the structure of the low frequency spectrum in the spin glass phase when $q_{EA} \ll 1$. 
For the spinon spectral density, we have
\beq
\rho (\omega)  = - \frac{1}{\pi} \mbox{Im} \, G_Q (\omega) = \frac{1}{\pi \sqrt{J \omega_\ast}} \Phi_\rho (\omega/\omega_\ast)\,, \label{eq:rho}
\eeq
with $\omega_\ast$ given by (\ref{omegastar}). 
The scaling function $\Phi_\rho$ obeys $\Phi_\rho (0) = 1$, and $\Phi_\rho (\overline{\omega} \gg 1) \sim 1/\sqrt{\overline{\omega}}$. We present result for $\rho(\omega)$ in Fig.~\ref{fig:spinon}, comparing with the scaling in (\ref{eq:rho}). 
\begin{figure}
\begin{center}
\includegraphics[width=4in]{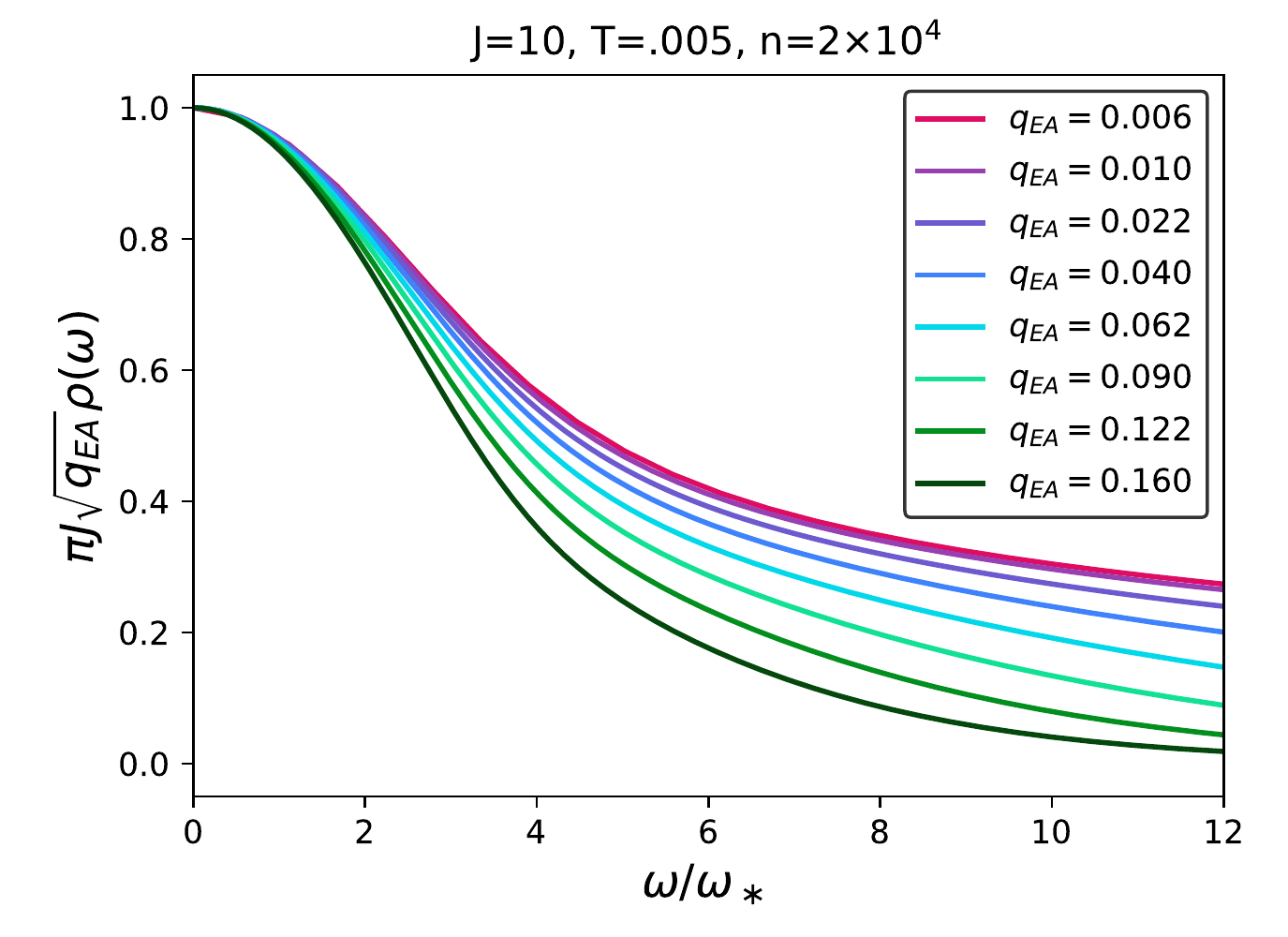}
\end{center}
\caption{Numerical results for the spinon spectral density obtained by the solution of (\ref{e1a}) and (\ref{e1bM}). The results scale as in (\ref{eq:rho}) for small $q_{EA}$. The solutions were obtained with $n$ frequency points.}
\label{fig:spinon}
\end{figure}

Similarly, for the spin spectral density we have
\beq
\chi'' (\omega) = - \mbox{Im} \, Q (\omega) = \frac{1}{J} \Phi_\chi (\omega/\omega_\ast)\,. \label{chiscale}
\eeq
Note that the full dynamic spin susceptibility has the delta function in (\ref{chidelta}), which is not included in  (\ref{chiscale}). The scaling function $\Phi_\chi (\overline{\omega})$ has the form given by (\ref{chilinear}) at $\overline{\omega} \ll 1$, and by (\ref{chisgn}) for $\overline{\omega} \gg 1$, and this is illustrated in Fig.~\ref{fig:chi}.
\begin{figure}
\begin{center}
\includegraphics[width=4in]{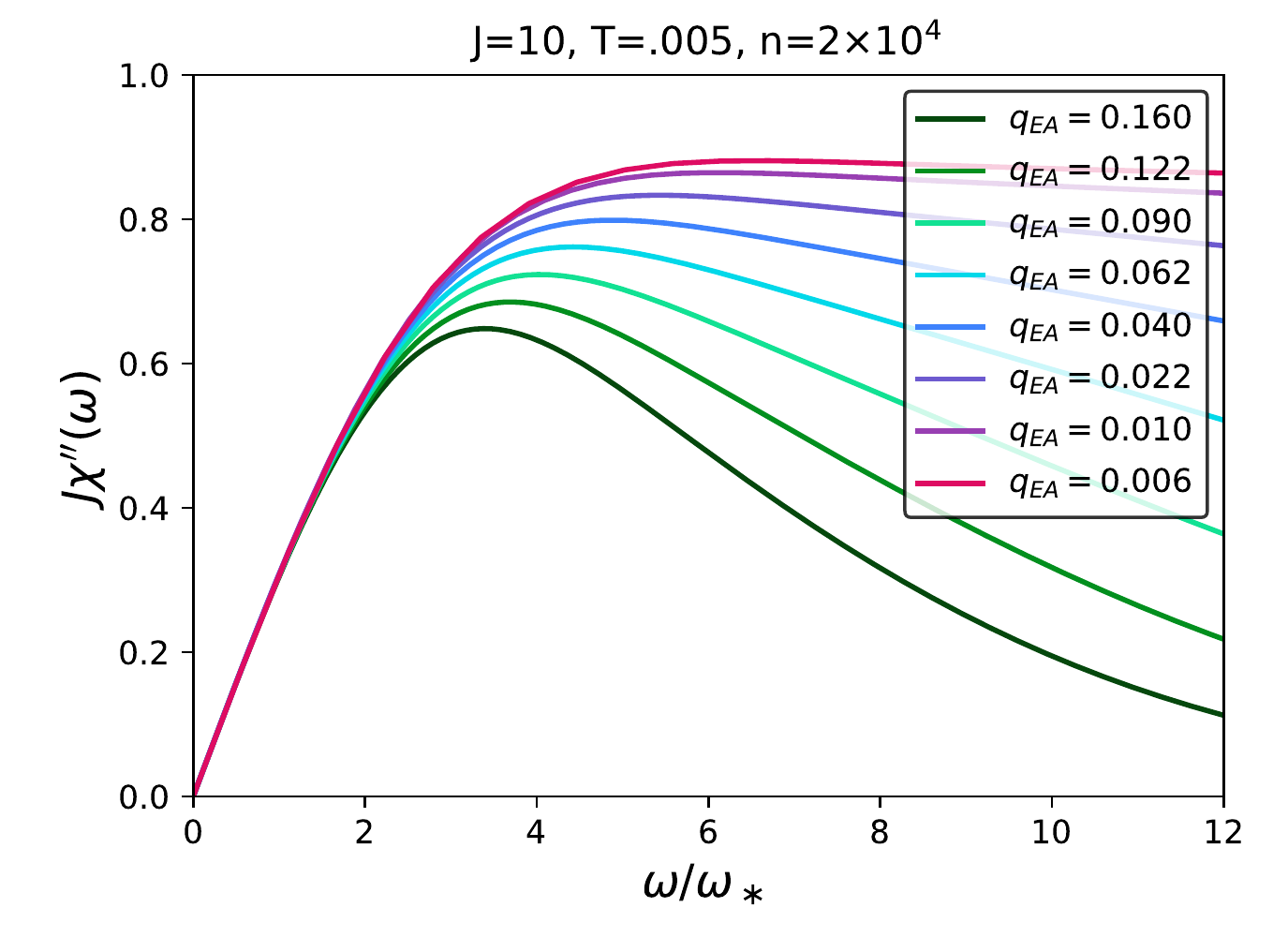}
\end{center}
\caption{Numerical results for the spin spectral density obtained by the solution of (\ref{e1a}) and (\ref{e1bM}). The results scale as in (\ref{chiscale}) for small $q_{EA}$.}
\label{fig:chi}
\end{figure}

The real part of the local spin response function {\it i.e.\/} the local static susceptibility has a logarithmic contribution which violates scaling: the $q_{EA} = 0$ result in (\ref{chiloc1}) is replaced by
\beq
\chi_{\rm loc} = \frac{1}{ J \sqrt{\pi}} \ln (J/\omega_\ast) \,, \quad T=0 \,. \label{chiloc2}
\eeq
This is illustrated in Fig.~\ref{fig:log}.
\begin{figure}
\begin{center}
\includegraphics[width=4in]{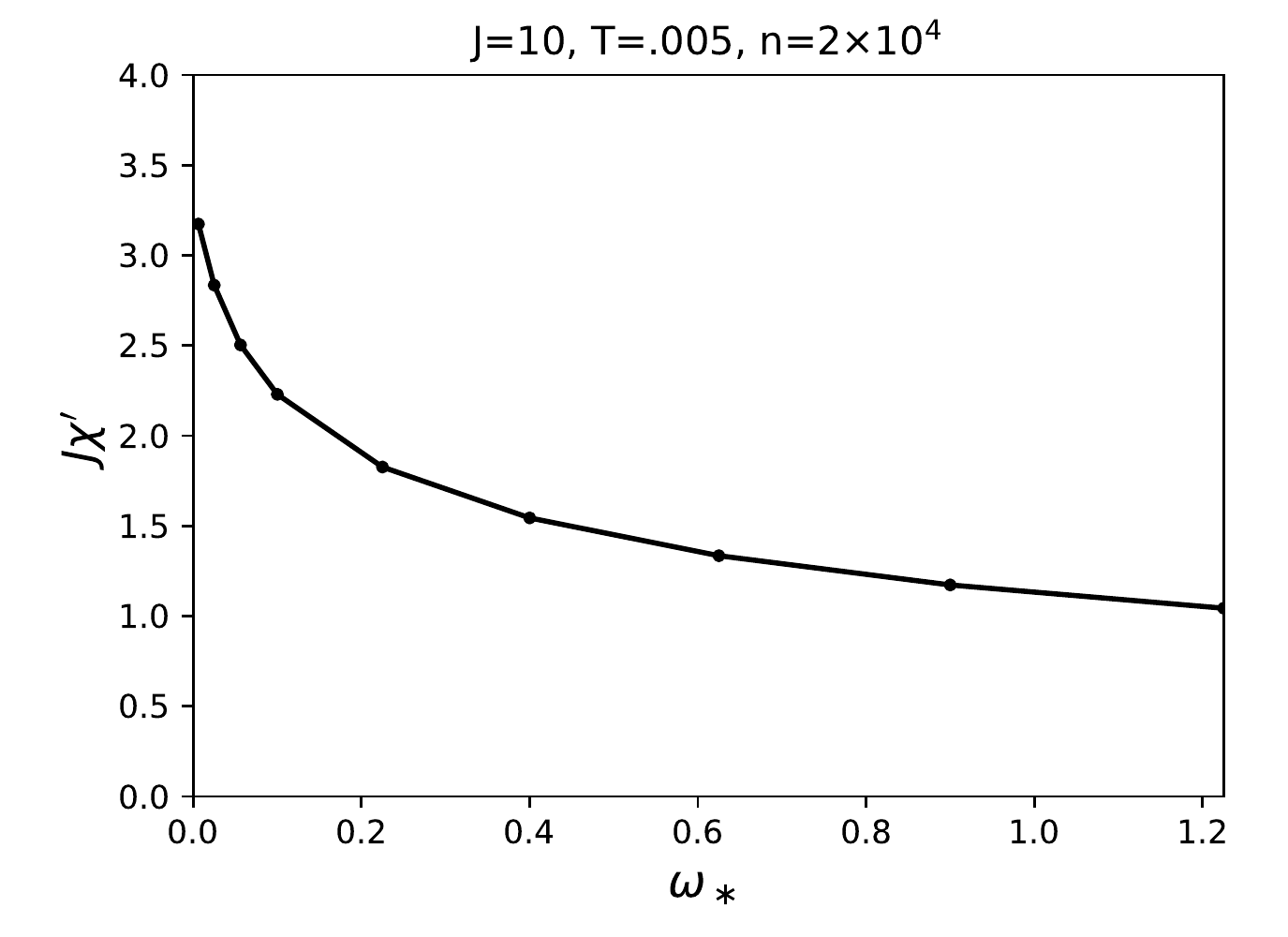}
\end{center}
\caption{The real part of the local susceptibility $\chi' = \chi_{\rm loc}$.}
\label{fig:log}
\end{figure}

\section{Complexity}
\label{sec:complexity}

In the spin liquid phase, the model features an exponential density of states and an extensive (in $N$) thermodynamic entropy.
As the system enters the spin glass phase and thermal fluctuations are further reduced, the thermodynamic entropy approaches zero. Instead, an extensive configurational entropy counts an exponential number of possible meta-stable glass states. This configurational entropy is often referred to as the complexity $\Sigma$, which can be expressed as a functional of the free energy ${\cal F}$ and the break-point parameter $m$ of the replica symmetry breaking ansatz \cite{monasson1995structural,Franz_1998,Mezard_1999}.\footnote{ See also \cite{zamponi2010mean} for a review, \cite{de2006random} for a textbook discussion, and \cite{Anous:2021eqj} for a recent application.} The number of meta-stable states in a given free energy band of width $1/m$ is then given by 
\begin{equation}
  \Omega({\cal F},m) = e^{N\Sigma({\cal F},m)} \,.
\end{equation}
In practice we compute the complexity as the Legendre transform of the free energy with respect to $m$. As a function of $(\beta,m)$, this can be computed as:
\begin{equation}
\label{eq:SigmaDefCalc}
 \Sigma(\beta,m)   = \beta m^2\, \partial_{m}  \,{\cal F}(\beta,m) \,.
 \end{equation}
We will now show how to evaluate this expression in our model. 

Consider the following simple Landau free energy, which exhibits the basic structure of our model:
\begin{equation}
\label{eq:Ftoy}
{\cal F}_{sg}[\overline{q},q_{ab}] = - d_2 \beta\,  \left(\overline{q}^2 + \frac{\text{Tr} q^2}{n} \right) - \frac{e_3}{3} \, \beta^2 \, \left( \overline{q}^3 +3\overline{q} \, \frac{\text{Tr}q^2}{n} + \frac{\text{Tr}q^3}{n} \right) -d_4 \beta \left( \overline{q}^4 + \frac{1}{n} \sum_{a ,b} q_{ab}^4 \right) + \ldots
\end{equation}
The coefficients $d_2$, $e_3$, $d_4$ are dimensionful but finite as $\beta \rightarrow \infty$. We computed $d_2$ and $d_4$ in Section \ref{sec:1M}; the coefficient $e_3$ is generated at ${\cal O}(M^{-1})$, see Appendix~\ref{app:higherOrders}. We dropped terms that do not contribute to the zero-temperature complexity, as well as terms at higher orders in $\overline q$ and $q_{ab}$.

To evaluate the free energy, we now go beyond the replica symmetric ansatz and consider full replica symmetry breaking (FRSB). This is implemented by starting from the Parisi ansatz for $k$-step replica symmetry breaking and then considering the limit $k\rightarrow \infty$: we first make an ansatz for $q_{ab}$ with constant blocks along the diagonal and then successively refine the structure by breaking up blocks into smaller blocks:
\begin{equation}
q_{ab}
=\left( \; \begin{array}{ccccc}
\cline{1-1} 
\multicolumn{1}{|c|}{A_{m_1}} & & & &  \\ \cline{1-2} 
 &  \multicolumn{1}{|c|}{A_{m_1}} &  & \;\; q_0 & \\ \cline{2-2}
 & & \ddots & & \\  \cline{4-4}
 & q_0 \;\; & & \multicolumn{1}{|c|}{A_{m_1}}    & \\  \cline{4-5}
 &  & &  & \multicolumn{1}{|c|}{A_{m_1}}    \\  \cline{5-5}
\end{array} \; \right)  \,,
\qquad
A_{m_1} =\left( \; \begin{array}{ccccc}
\cline{1-1} 
\multicolumn{1}{|c|}{A_{m_2}} & & & &  \\ \cline{1-2} 
 &  \multicolumn{1}{|c|}{A_{m_2}} &  & \;\; q_1 & \\ \cline{2-2}
 & & \ddots & & \\  \cline{4-4}
 & q_1 \;\; & & \multicolumn{1}{|c|}{A_{m_2}}    & \\  \cline{4-5}
 &  & &  & \multicolumn{1}{|c|}{A_{m_2}}    \\  \cline{5-5}
\end{array} \; \right) 
\end{equation}
and so on, where $A_{m_i}$ is an $m_i\times m_i$ matrix with blocks $A_{m_{i+1}}$ along the diagonal and all off-diagonal entries filled with $q_{i}$ such that $n = \sum_i m_i$.
In the analytic continuation $n\rightarrow 0$ we replace the matrix $q_{ab}$ by a monotonously increasing function $q(x)$, which extrapolates the structure above. The variable $x\in [0,1]$ parametrizes continuous breaking of replica symmetry. The analog of the Edwards-Anderson parameter is $q_{EA} = q(x=1)$. 

With this ansatz, the free energy \eqref{eq:Ftoy} then becomes a functional of $q(x)$:
\begin{equation}
\begin{split}
{\cal F}_{sg}[\overline{q},q(x)] &=  \int_0^1 dx \, \bigg\{ d_2\beta  \left(   q(x)^2-\overline{q}^2 \right) + d_4\beta \, \left( q(x)^4- \overline{q}^4 \right) \\
 &\qquad \qquad\quad - \frac{e_3}{3}\, \beta^2 \,  \left( \overline{q}^3 - 3 \overline{q} \, q(x)^2  + x q(x)^3 + 3 q(x) \int_0^x dy \, q(y)^2 \right) + \ldots\bigg\}
\end{split}
\end{equation}
Extremizing the action with respect to $q(x)$ leads to the following continuous solution \cite{fischer1993spin}:
\begin{equation}
q(x) = \left\{ \begin{aligned} &\frac{x}{m} \, q_{EA}\,, \qquad\; x\in \left[ 0, m\right]\\
  &q_{EA}\,, \qquad\quad\;\; x \in \left[ m, 1\right]
  \end{aligned} \right.
\end{equation}
where $m  \equiv \frac{12 d_4}{e_3\beta} \,q_{EA}$ plays the role of the break point parameter of an equilibrium solution. The extremization procedure also relates the value of the diagonal contribution $\overline{q}$ to $q_{EA}$: 
\begin{equation}
\label{eq:mSolSimpl}
 \overline{q} = q_{EA} - \frac{d_2 + 6 d_4 q_{EA}^2}{e_3 \beta} \,.
\end{equation}
Note that we do not separately extremize with respect to $\overline{q}$ because even perturbatively the ansatz \eqref{eq:Ftoy} only captures part of the full $\overline{q}$-dependence of our system.

Evaluated on the saddle point solution for $q(x)$, the free energy takes the following value:
\begin{equation}
\begin{split}
{\cal F}_{sg}(m,\beta_e) &= - \frac{d_2^4 d_4 m^3}{e_3^4 \beta_e^3} - \frac{d_2^3 m(4-2m+m^2)}{6e_3^2 \beta_e} + \frac{d_2^2(2-m)(4-2m+m^2)\beta_e}{96 d_4} \\
&\quad\, + \frac{d_2 e_3^2 (2-m)^3 \beta_e^3}{6 (24 d_4)^2} + \frac{e_3^4[48-5m(4-m)^2]\beta_e^5}{15 (48d_4)^3} + \ldots\,,
\end{split}
\end{equation}
where $\beta_e \equiv m\beta$ is an effective temperature, conjugate to the free energy. Expanding in large $\beta$, we obtain:
\begin{equation}
\begin{split}
{\cal F}_{sg}(m,\beta)  &= \frac{d_2^2}{e_3} \, q_{EA} + \frac{4d_2 d_4}{e_3} \, q_{EA}^3 + \frac{36d_4^2}{5e_3} \, q_{EA}^5 + {\cal O}(\beta^{-1}) + \ldots 
  \end{split}
\end{equation}
Note again that the free energy is finite as $\beta \rightarrow \infty$. This was not guaranteed to happen. It is a consequence of the specific way in which $\overline{q}$ appeared in \eqref{eq:Ftoy} and of the extremization condition \eqref{eq:mSolSimpl}.

In order to compute the low temperature complexity we take an $m$-derivative (at fixed $\beta$) and then expand in large $\beta$. According to \eqref{eq:SigmaDefCalc}, we find:
\begin{equation}
\begin{split}
 \Sigma(m,\beta)& \equiv \beta m^2 \partial_m  {\cal F}_{sg}(m,\beta)   \equiv \beta_e^2 \partial_{\beta_e} {\cal F}_{sg}(m,\beta_e)\\
 & = \frac{12 d_4}{e_3^2}\, q_{EA}^2 \left( d_2  + 6d_4 q_{EA}^2 + \ldots  \right)^2 + {\cal O}(\beta^{-1}) \,.
\end{split}
\end{equation}
The fact that the complexity $\Sigma$ is finite as $\beta \rightarrow \infty$ means that the spin glass at zero temperature is characterized by an extensive number $e^{N\Sigma}$ of meta-stable states.

\section{Discussion}

The initial analysis \cite{SY92} of the SU($M$) random quantum magnet (\ref{SUMHam}) found a gapless spin liquid ground state in the large $M$ limits realized by fermionic and bosonic spinons, and both limits yielded a `marginal' dynamic spin susceptibility with $\chi'' (\omega) \sim \mbox{sgn} (\omega)$ at small $\omega$. 
This fractionalized spin liquid is unstable to spin glass order at low enough temperatures for any finite $M$ \cite{GPS01}, and (\ref{Tc}) contains an estimate of the critical temperature for fermionic spinons.
A theory of a spin glass ground state was presented in Refs.~\cite{GPS00,GPS01} using bosonic spinons, in which case the spin glass order can be large, with $q_{EA} = \mathcal{O}(M^0)$ (see (\ref{maxqb})). However, numerical studies of the SU(2) case show that the spin glass order is small \cite{ArracheaRozenbergSG2002,Shackleton2021}, and the intermediate frequency spin spectrum was a better match with the large $M$ theory with fermionic spinons \cite{Shackleton2021,Tikhanovskaya:2020elb}. Here we have presented an analysis which is closest to the numerical observations: a theory for the onset of weak spin glass order using fermionic spinons, where $q_{EA}$ is at most $\mathcal{O}(M^{-1})$ (see (\ref{maxq})). We identified a frequency scale $\omega_\ast = J q_{EA}$, and showed that $\chi'' (\omega) \sim \omega$ in the fermionic spinon theory, as had also been found for small $\omega$ in the bosonic spinon theory. For the case of small $q_{EA}$, there is a universal crossover from the physics of a spin liquid with fractionalized spinons for $\omega > \omega_\ast$, to the physics of a confining spin glass for $\omega < \omega_\ast$, and we obtained results for the crossover functions.

In Section~\ref{sec:spectrum}, we mapped the crossover from the spectrum of the SYK spin liquid to the spin glass to the crossover from non-Fermi liquid to Fermi liquid behavior in the model of Ref.~\cite{Balents2017}.
We now comment on why this can be interpreted as a crossover from fractionalization to confinement in our context of the random quantum magnet. Unlike the case for the model of Ref.~\cite{Balents2017}, the fermions in our quantum magnet, and in the $t$-$J$ models of Refs.~\cite{PG98,randomtJ1}, carry a U(1) gauge charge: (\ref{SkM}) is invariant under the gauge transformation $f^\alpha (i) \rightarrow f^\alpha (i) e^{i \phi_i (\tau)}$ (see Appendix~\ref{app:U1}). Consequently the SYK spin liquid can be regarded as a gapless spin liquid with fractionalized fermionic spinons. The crossover to the spin liquid phase is induced by the $q_{EA}$ term in (\ref{e1bM}), which turns out to be identical to the influence of the $t$ term in the $t$-$J$ models of Refs.~\cite{PG98,randomtJ1}. The latter $t$ term is known to break the U(1) gauge symmetry, and therefore, by Higgs-confinement continuity, we can regard the low frequency regime of our quantum magnet as a confining regime of the U(1) gauge symmetry. It is also interesting to compare with the analysis of the quantum magnet using bosonic spinons in Refs.~\cite{GPS00,GPS01}: that model also exhibits a fractionalized spin liquid regime, and spin glass order appears by the condensation of bosonic spinons, which explicitly higgses the U(1) gauge symmetry (see Appendix~\ref{app:bosons}). 
Moreover the dynamic spectrum $\chi'' (\omega) \sim \omega$ appears not only for bosonic and fermionic spinons in the spin glass regime, but also for the Ising and rotor spin glasses \cite{RSY95,Cugliandolo00,Anous:2021eqj} where there is no fractionalization at any frequency scale.
So, as we noted in Section~\ref{sec:intro}, the random quantum magnet analyzed here yields a realization of fermion-boson duality, and a solvable theory of deconfinement-confinement crossover in a gapless system with finite density matter. We are not aware of other solvable examples of such phenomena.

In Section~\ref{sec:complexity}, we employed the insights gained from the structure of the 
spin glass state to make some general remarks on the complexity of infinite-range quantum spin glasses in the low temperature limit. Our main result was that the complexity is generically non-zero and extensive in the limit of vanishing temperature. For the random quantum magnets considered here, the SU($M \rightarrow \infty$) models have a spin liquid ground state with a non-zero extensive entropy in the limit of vanishing temperature \cite{GPS01} (here `extensive' refers to proportionality to $N$, the number of sites, and not to $M$). For finite $M$, we have shown that this entropy is quenched at an energy scale $\omega_\ast$. Below $\omega_\ast$, 
we obtain a spin glass state which in the limit of vanishing temperature has no extensive entropy but an extensive complexity. It appears that the chaotic quantum dynamics in the exponentially large phase space explored by the spin liquid gets turned off at low temperatures, and the phase space fragments into an exponentially large number of subspaces.
It would be interesting to explore this idea in the context of the holographic nAdS$_2$/nCFT$_1$ paradigm, which gives a gravitational interpretation of the low-energy Schwarzian sector describing the spin liquid phase at strong coupling \cite{kitaev2015talk,Maldacena:2016upp}: motivated by the existence of landscapes of multi-centered black hole solutions in four dimensional supergravity \cite{Denef:2000nb,Bates:2003vx}, it was previously suggested \cite{Anninos:2011vn,Anninos:2013mfa,Anninos:2016szt,Anous:2021eqj} that the spin glass crossover could be realized gravitationally in terms of the fragmentation instability of AdS${}_2$ spacetimes \cite{Maldacena:1998uz}. The latter gives rise to a landscape of asymptotically AdS${}_2$ geometries characterized by the number, location, and charge of fragmented throats. It might then be possible to interpret the complexity of the spin glass state as a measure of the volume of the moduli space of gravitational solutions.

\subsection*{Acknowledgements}

We thank Tom Banks, Debanjan Chowdhury, Antoine Georges, Darshan Joshi, Chenyuan Li, Juan Maldacena, Olivier Parcollet, Henry Shackleton, Grigory Tarnopolsky, Maria Tikhanovskaya, and Alexander Wietek for helpful discussions.
This research was supported by the National Science Foundation under Grant No.~DMR-2002850. This work was also supported by the Simons Collaboration on Ultra-Quantum Matter, which is a grant from the Simons Foundation (651440, S.S.). F.H.\ is supported by the U.S.\ Department of Energy, Office of Science, Office of High Energy Physics under Award Number DE-SC0009988, and by the Paul Dirac and Sivian Funds.

\appendix

\section{Bosonic spinons}
\label{app:bosons}

The Appendix briefly reviews the bosonic spinon theory of the spin glass state \cite{GPS00,GPS01} of (\ref{SUMHam}).

Each site now contains states corresponding to the {\it symmetric\/} product of $\kappa M$ (integer) fundamentals, and (\ref{SkM}) is replaced by
\beq
\mathcal{S}_{\beta}^\alpha (i) = b_\beta^\dagger (i) b^\alpha (i) - \kappa \delta^{\alpha}_{\beta} \,,\quad \sum_\alpha  b_\alpha^\dagger (i) b^\alpha (i) = \kappa M \label{SkMb}
\eeq
with bosons $b^\alpha (i)$ on each site $i$.
The bosonic and fermionic spinon models co-incide only for the SU(2) case of physical interest, with $M=2$, $\kappa = k = 1/2$.

Now the perfectly ordered spin-glass has $\kappa M$ bosons in the $\alpha = 1$ state (say), and this replaces the bound in (\ref{maxq}) by
\beq
q_{EA} \leq \frac{\kappa^2 (M-1)}{M} \,. \label{maxqb}
\eeq
Note that (\ref{maxq}) and (\ref{maxqb}) agree for the SU(2) case. However, unlike (\ref{maxq}), the bound in (\ref{maxqb}) does not vanish in the $M \rightarrow \infty$ limit, and so spin glass order can be order unity in $M=\infty$ theory. This order is realized by a Higgs condensate of the bosonic spinons \cite{GPS00,GPS01}
\beq
\left\langle b^\alpha \right\rangle =  \sqrt{M} \left( q_{EA} \right)^{1/4} \delta_{\alpha,1} \,.
\eeq
This condensate breaks the U(1) gauge symmetry associated with the bosonic analog of (\ref{SkMb}). In the replica theory, condensate requires replica off-diagonal components in the boson Green's function $G_ab$ at zero frequency \cite{GPS00,GPS01}
\beq
G_{ab} (i \omega_n) = \delta_{ab} G(i \omega_n) + \beta \delta_{\omega_n, 0} \,g_{ab}\,.
\eeq
The replica off-diagonal components of $g_{ab}$ break replica symmetry, and this symmetry breaking has to satisfy a marginal stability criterion to obtain a gapless boson spectrum.

\section{U(1) gauge invariance}
\label{app:U1}

Consider the following gauge transformation:
\begin{equation}
    \begin{split}
    f_a^\alpha(\tau) &\longrightarrow e^{i\phi(\tau)}  f_a^\alpha(\tau) \,, \qquad 
    f^\dagger_{a\alpha}(\tau) \longrightarrow e^{-i\phi(\tau)}  f^\dagger_{a\alpha}(\tau) \,, \qquad
    \lambda_a(\tau) \longrightarrow \lambda_a(\tau) - \partial_\tau \phi \,.
    \end{split}
\end{equation}
This is an invariance of the fermionic formulation of the theory, e.g., \eqref{SQ}. In the $G$-$\Sigma$ formulation \eqref{ZfQ}, one can see that the action is invariant under the following transformations:
\begin{equation}
    \begin{split}
    G_{ab}(\tau,\tau') &\longrightarrow e^{i[\phi(\tau)-\phi(\tau')]}\, G_{ab}(\tau,\tau')\\
    \Sigma_{ab}(\tau,\tau') &\longrightarrow e^{i[\phi(\tau)-\phi(\tau')]}\, \Sigma_{ab}(\tau,\tau')\\
    \lambda_a(\tau) & \longrightarrow \lambda_a(\tau) - \partial_\tau \phi\,.
    \end{split}
\end{equation}
We can use this gauge symmetry to make $\lambda_a$ time-independent, but cannot remove it entirely because of the periodicity condition on the fields.

\section{Higher orders in $1/M$}
\label{app:higherOrders}

In this appendix we compute higher orders in the $1/M$ expansion of the free energy.

We first clarify some notation: we use a matrix dot product both for fields with two and with four indices. Every `matrix' multiplication always involves half of the available indices. Relevant quantities occurring below are:
\begin{equation}
    \begin{split}
    \delta {\bf \Sigma} \equiv \delta \Sigma_{ab}(\tau_1,\tau_2) \,,\quad {\bf G_Q} \equiv G_Q(\tau_{12}) \delta_{ab} \,,\quad {\bf K} \equiv K_{ab;cd}(\tau_1,\tau_2;\tau_3,\tau_4) \,.
    \end{split}
\end{equation}
These multiply as follows:
\begin{equation}
    \begin{split}
    \delta {\bf \Sigma} \cdot {\bf G_Q} &= \int d\tau_3 \sum_c \delta \Sigma_{ac}(\tau_1,\tau_3) (G_Q(\tau_{32}) \delta_{cb} )\,, \\
    {\bf K} \cdot {\bf K} &= \int d\tau_5 d\tau_6 \sum_{ef} K_{ab;ef}(\tau_1,\tau_2;\tau_5,\tau_6) K_{ef;cd}(\tau_5,\tau_6;\tau_3,\tau_4) \,.
    \end{split}
\end{equation}

Let us now explain the $1/M$ expansion of the free energy. We need to consider higher powers of $\delta \Sigma_{ab}$ in the expansion of $I[Q]$. From \eqref{ZfQ}, we find that such terms originate from expanding the logarithm: 
\begin{equation}
\begin{split}
& - \ln \det \left\{- \delta'(\tau-\tau')  \delta_{ab} - (\Sigma_{ab}(\tau,\tau')+\delta\Sigma_{ab}(\tau,\tau')) \right\}  \\
&\qquad = \ldots + \frac{1}{3} \, \text{Tr} \left( \delta {\bf \Sigma}\cdot{\bf G_Q}  \right)^3 + \frac{1}{4} \, \text{Tr} \left(   \delta {\bf \Sigma}\cdot {\bf G_Q}\right)^4 + {\cal O}(\delta {\bf\Sigma}^5) \,.
\end{split}
\end{equation}
where we omitted constant, linear and quadratic terms, which are already taken care of.
In the functional integral over $(\delta G_{ab}, \delta\Sigma_{ab})$ we include these higher order terms by introducing a bilocal source $J_{ab}$ for $\delta \Sigma_{ab}$ in the integral \eqref{eq:fluctInt}:
\begin{equation}
\label{eq:fluctIntPert}
\begin{split}
{\cal Z}_f[Q] &\propto \exp \left[- \sum_{k\geq 3} \frac{M}{k}  \, \text{Tr} \left(\frac{\delta}{\delta {\bf J}} \right)^k  \right]_{{\bf J}=0}\, \int {\cal D}[\delta {\bf X}] \, e^{ -\frac{M}{2} \delta {\bf X}^\text{T} \cdot {\bf A} \cdot \delta {\bf X} + \text{Tr}[ {\bf J}\cdot \delta {\bf \Sigma} \cdot {\bf G_Q} ]} \\
&\propto [\det ({\bf K}-{\bf 1})]^{-\frac{1}{2}}\,  \exp \left[- \sum_{k\geq 3} \frac{M}{k} \, \text{Tr} \left(  \frac{\delta}{\delta {\bf J}} \right)^{k}  \right]_{{\bf J}=0} \,
\exp \left[ \frac{1}{M} \, W_Q[{\bf J}] \right] \,,
\end{split}
 \end{equation}
where we discarded the leading contribution obtained by evaluating ${\cal Z}_f$ on the large $M$ saddle point, and we defined
\begin{equation}
     W_Q[{\bf J}] \equiv -\frac{J^2}{2} \int  d\tau_1 \cdots d\tau_6 \sum_{a,b,c,d} J_{ab}(\tau_1,\tau_2) J_{cd}(\tau_3,\tau_4) \, Q_{cd}(\tau_6,\tau_4) G_Q(\tau_{15}) G_Q(\tau_{36}) \,({\bf 1}- {\bf K})^{-1}_{ab;cd}(\tau_2,\tau_5;\tau_6,\tau_4)
\end{equation}

In order to compute the subleading contributions to the free energy, we need to evaluate the new contributions to $-\ln {\cal Z}_f[Q]$, which are generated by derivatives with respect to ${\bf J}$. Note that the logarithm does not simply remove the exponential in \eqref{eq:fluctIntPert} due to the structure of contractions. For instance, the terms involving four and six ${\bf J}$-derivatives take the following form:
\begin{equation}
\label{eq:fluctIntPert2b}
\begin{split}
-\ln {\cal Z}_f[Q] 
&= \ldots  + \left\{ \frac{1}{8M} \, \text{Tr} \left(  \frac{\delta}{\delta {\bf J}} \right)^4 W^2 +  \left( \frac{1}{36M^2} \, \text{Tr} \left(  \frac{\delta}{\delta {\bf J}}\right)^6
- \frac{1}{108 M} \, \left[\text{Tr} \left(  \frac{\delta}{\delta {\bf J}} \right)^3 \right]^2 \right) W^3 + \ldots \right\}
\end{split}
 \end{equation}

It is most useful to think about these expressions diagrammatically: the ${\bf J}$-derivatives produce different Wick contractions among the powers of ${\bf K}$.
For instance, at ${\cal O}(M^{-1})$ we obtain the following contribution to the free energy from the first term in \eqref{eq:fluctIntPert2b}:
\begin{equation}
\label{eq:higherCorr1}
 \begin{split} 
\includegraphics[width=0.085\textwidth, valign=c]{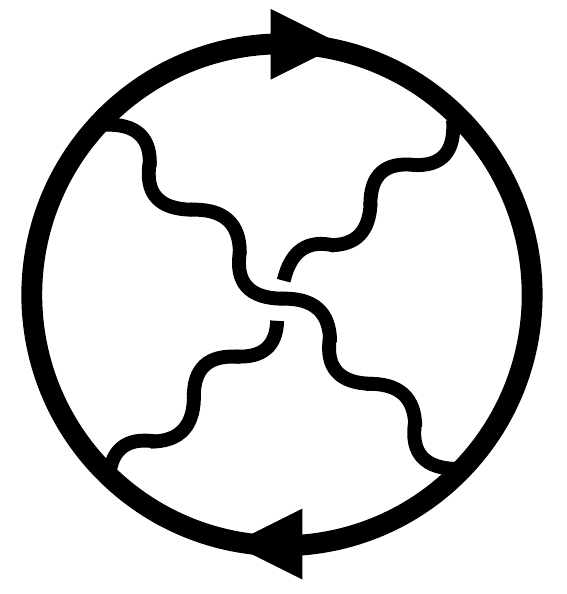} 
  &\propto \frac{J^4}{M} \int d\tau_1 d\tau_2d\tau_3d\tau_4 \, \sum_a Q_{aa}(\tau_{13}) Q_{aa}(\tau_{24}) R_Q^{(4)}(\tau_{14},\tau_{43},\tau_{32}) 
\end{split}
\end{equation}  
Similarly, the last term shown in \eqref{eq:fluctIntPert2b} gives further contributions at ${\cal O}(M^{-1})$, such as:
\begin{equation}
\label{eq:cubic1}
\begin{split} 
\includegraphics[width=0.098\textwidth, valign=c]{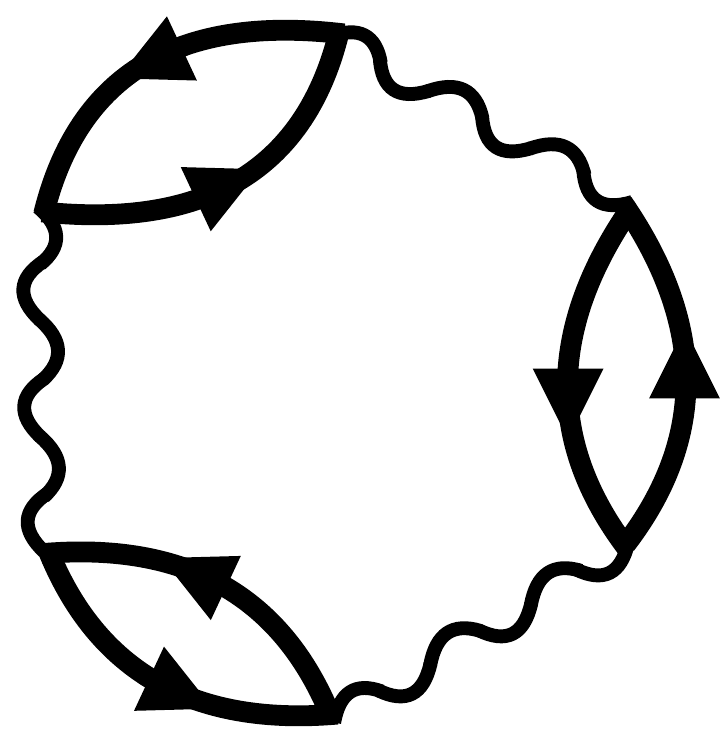} &\propto  \frac{J^6}{M} \int d\tau_1\cdots d\tau_6 \sum_{a,b,c}  Q_{ab}(\tau_{12}) Q_{bc}(\tau_{34})Q_{ca}(\tau_{56}) R_Q^{(2)}(\tau_{23}) R_Q^{(2)}(\tau_{45}) R_Q^{(2)}(\tau_{61})
\end{split}
\end{equation} 
At ${\cal O}(M^{-2})$ we get cubic terms such as the following from the second term shown in \eqref{eq:fluctIntPert2b}:
\begin{equation}
\label{eq:cubicDiags}
 \begin{split} 
\includegraphics[width=0.118\textwidth, valign=c]{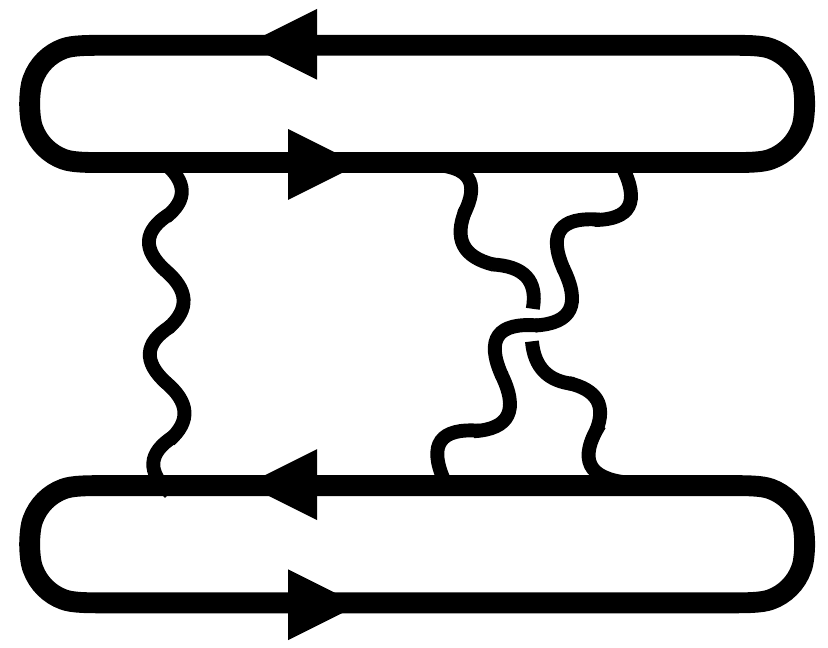} 
  &\propto \frac{J^6}{M^2} \int d\tau_1 \cdots d\tau_6 \sum_{a,b} Q_{ab}(\tau_{14}) Q_{ab}(\tau_{25}) Q_{ab}(\tau_{36}) R_Q^{(3)}(\tau_{15},\tau_{53}) R_Q^{(3)}(\tau_{42},\tau_{26})\\
\includegraphics[width=0.118\textwidth, valign=c]{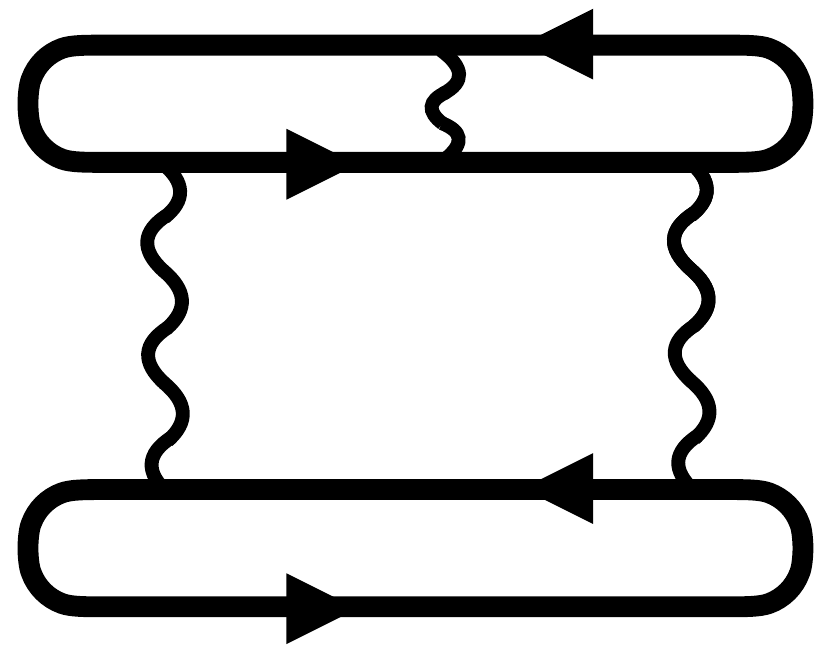} 
  &\propto \frac{J^6}{M^2} \int d\tau_1 \cdots d\tau_6 \sum_{a,b}  Q_{ab}(\tau_{13}) Q_{aa}(\tau_{25}) Q_{ab}(\tau_{46}) R_Q^{(2)}(\tau_{14}) R_Q^{(3)}(\tau_{32},\tau_{26},\tau_{65})
\end{split}
\end{equation} 

We can now see how further potentially divergent terms are generated in the free energy functional at higher orders in $1/M$. For example, the diagram \eqref{eq:cubic1} and the last diagram shown in \eqref{eq:cubicDiags} lead to new contributions to the free energy, which are cubic in the spin glass parameters:
\begin{equation}
     {\cal F}_{sg} \supset  - \frac{e_3}{3} \, \beta^2  \left( \overline{q}^3 +3\overline{q} \, \frac{\text{Tr}q^2}{n}+\frac{\text{Tr}q^3}{n} \right)
     - e_3' \, \beta \, \overline{q} \left( \overline{q}^2 + \frac{\text{Tr}q^2}{n} \right) + \ldots
\end{equation}
where $e_3 = {\cal O}(M^{-1})$ and $e_3' = {\cal O}(M^{-2})$.
All possible terms in the Landau functional theory (e.g., Ref.\ \cite{RSY95}) are generated systematically this way.
The other diagrams shown above give $1/M$ corrections to the coefficients $c_i$ and $d_i$ that we already included in \eqref{eq:Fsg}.

Note that the free energy contribution proportional to $e_3$ is naively quadratically divergent as $\beta \rightarrow \infty$. However, upon using the replica symmetric ansatz for $q_{ab}$ and the extremization condition \eqref{eq:qbarqEA}, this divergence is again cured and we obtain a finite limit. This follows from the identity 
\begin{equation}
    \overline{q}^3 +3\overline{q} \, \frac{\text{Tr}q^2}{n}+\frac{\text{Tr}q^3}{n}  
    \;\; \longrightarrow \;\; \overline{q}^3 -3 \overline{q} \, q_{EA}^2 +2 q_{EA}^3
\end{equation}
for the replica symmetric ansatz as $n \rightarrow 0$.

\bibliography{sg}

\end{document}